\documentclass[epj]{webofc}
\usepackage[varg]{txfonts}   
\wocname{EPJ Web of Conferences}
\woctitle{CONF12}
\def \als {\alpha_{\mathrm{s}}}

\newcommand\lQ{\Lambda_{QCD}}
\newcommand{\be}{\begin{equation}}
\newcommand{\ee}{\end{equation}}
\newcommand{\bea}{\begin{eqnarray}}
\newcommand{\eea}{\end{eqnarray}}
\newcommand{\nn}{\nonumber}

\begin{document}
\selectlanguage{english}
\title{Heavy Hybrids: decay to and mixing with Heavy Quarkonium}
%
%

\author{Rub\'en Oncala\inst{1,2} \and
        Joan Soto\inst{1}\fnsep\thanks{\email{joan.soto@ub.edu}} 
}

\institute{Departament de F\'\i sica Qu\`antica i Astrof\'\i sica and Institut de Ci\`encies del Cosmos, Universitat de Barcelona, Mart\'\i $\,$ i Franqu\`es 1, 08028 Barcelona, Catalonia, Spain.
\and
  Nikhef, Science Park 105, 1098 XG Amsterdam, The Netherlands.
}

\hfill ICCUB-16-037
\newline

\hfill NIKHF-2016-054

\abstract{%
  We report on a recent QCD based research on hybrid mesons containing $c\bar c$ or $b\bar b$ quarks. We present results for the 
	spectrum, the decay widths to heavy quarkonium, and the role of mixing with the latter. We point out that mixing with heavy quarkonium provides a potentially large source of spin symmetry breaking. We identify candidates to hybrid mesons among the so called XYZ states in the charmonium and bottomonium spectrum. 
}
\maketitle
\section{Introduction}
\label{intro}
\begin{figure}[h]
\centering
\includegraphics[width=10cm,height=10cm]{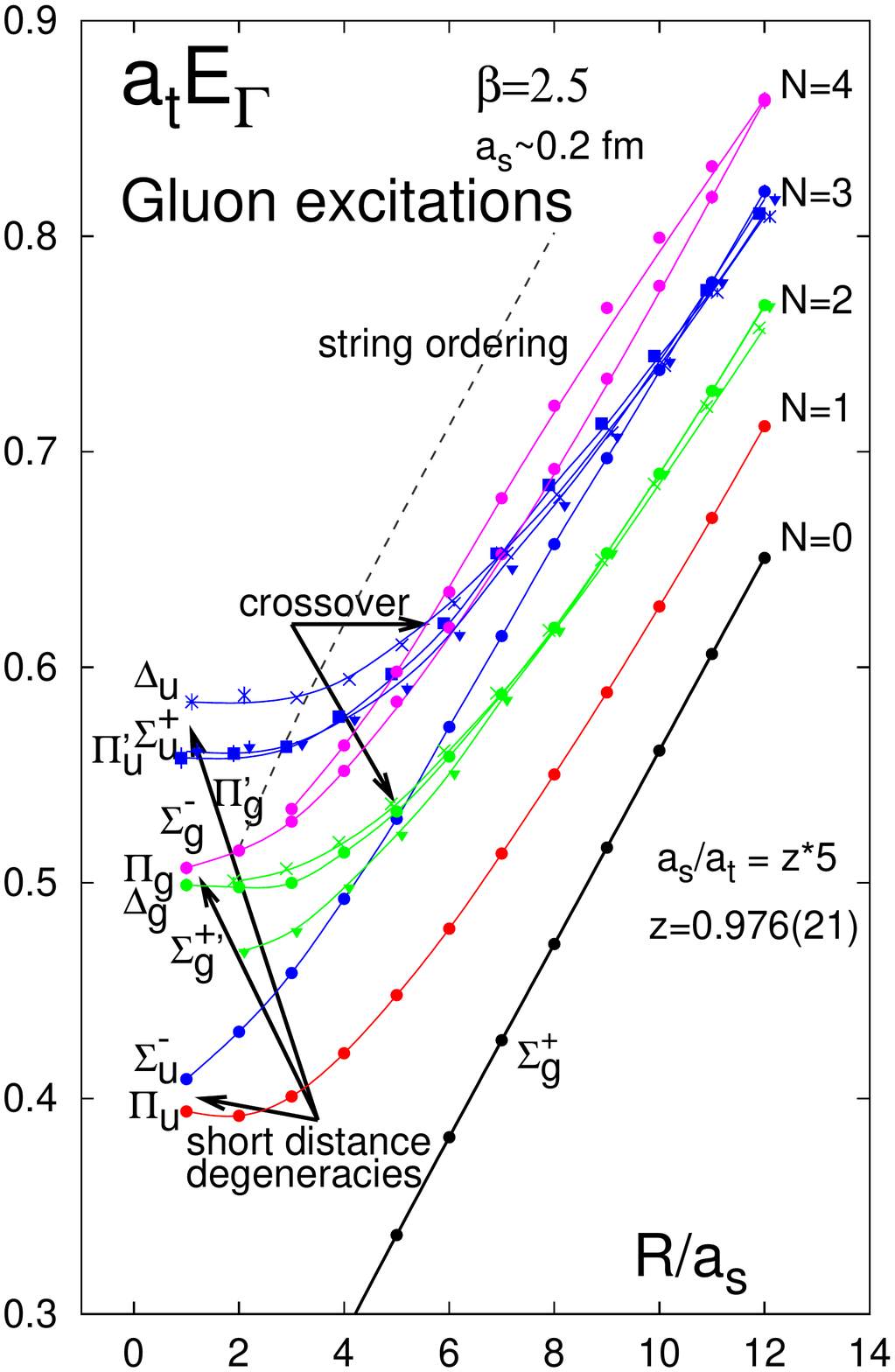}
\caption{Energy spectrum in the static limit for $n_f=0$ \cite{Juge:2002br}.}
\label{fig2}       
\end{figure}

The so called XYZ states in the charmonium and bottomonium spectrum do not fit in the usual 
potential model expectations (see \cite{Olsen:2015zcy} for a recent review). A number of models 
have been proposed to understand them, ranging from compact 
tetraquark states to just kinematical enhancements caused by the heavy-light meson pair 
thresholds. 
We pursue here a QCD based approach based on the fact that charm and bottom masses are much 
larger than the typical QCD scale $\lQ$, and hence Non-Relativistic QCD (NRQCD) \cite{Caswell:1985ui}
holds for these
 states. Furthermore, if we focus on a region of the spectrum much smaller than $\lQ$, we should 
be able to build an effective theory in that region, by integrating out $\lQ$, in a way similar to the 
strong coupling regime of Potential NRQCD (pNRQCD) \cite{Brambilla:1999xf}. The static limit is relevant for such a 
construction and the spectrum in that limit is known from lattice QCD in the case of $n_f=0$ 
(no light quarks) \cite{Juge:2002br}.
 In the Born-Oppenheimer (BO) approximation,
each energy level in the static case plays the role of a potential in a Schr\"odinger equation for
 the dynamical states build on that static energy level \cite{Perantonis:1990dy}. The static spectrum is displayed
 in fig. \ref{fig2}. 

The ground state corresponds to the potential for heavy quarkonium states, namely the 
one that it is usually input in potential models. The higher levels correspond to gluonic 
excitations and are called hybrid potentials. If we are interested in states of a certain energy,
 we must in principle take into account all
the potentials below that energy, since the states build on different potentials may influence each 
other through $1/m_Q$ corrections, $m_Q$ being the mass of the heavy quarks ($Q=c,b$).
We shall focus here on the lower lying hybrid states build out of $\Pi_u$ and $\Sigma_u^-$, and will 
address the question on how they talk to states build out of $\Sigma_g^+$ (heavy quarkonium). 
The $\Sigma_g^+$ states far below the energy of the $\Pi_u$ and $\Sigma_u^-$ can be integrated 
out and may contribute to the decay width, whereas the $\Sigma_g^+$ states in the same energy range as
 $\Pi_u$ and $\Sigma_u^-$ may mix with them. We shall address these issues in Sections \ref{decay} 
and \ref{mixing} respectively.
In Section \ref{spectrum} we calculate the spectrum of the $\Pi_u$ and $\Sigma_u^-$ states al 
leading order.

Before closing the introduction, let us briefly discuss the important question on how fig. \ref{fig2}
 changes in the case $n_f=3$ (three light quarks). 
We know that $\Sigma_g^+$ does not change
much and this is also so for $\Pi_u$,  at least up to moderately large distances \cite{Bali:2000un}.
 Nothing is known about the rest, but there is no reason 
to expect a different behavior. Two major qualitative features arise though. The first one is 
the appearance 
of heavy-light meson pairs, which amount to roughly horizontal lines at the threshold energies. 
These states talk to the remaining potentials already at leading order, and may in principle 
produce important distortions with respect to the $n_f=0$ case.
In practice, we only know how they cross talk to the $\Sigma_g^+$ state, and turn out to produce a tiny
 disturbance to the spectrum, apart from avoiding level crossing \cite{Bali:2005fu}. Hence we 
expect the effects of
 $n_f\not=0$ to be important only when our states are very close to some heavy-light meson pair 
threshold. 
The second one is the appearance of light quark excitations, in addition to the gluon ones, in 
the 
static spectrum. They may have different quantum numbers, for instance nonzero isospin (in this
case they may be relevant to the experimentally discovered charged $Z$ states). We do not know 
anything about those and, as pointed out in \cite{Brambilla:2008zz} 
and more recently emphasized in \cite{Braaten:2013boa,Braaten:2014qka}, it would be extremely 
important to have lattice QCD evaluations of the static energies of light quark excitations. For the time being, we shall neglect the mixing with those states.

\section{Spectrum}
\label{spectrum}
In the BO approximation, the calculation of the heavy hybrid spectrum reduces to solving the Schr\"odinger equation
with a potential $V=V(r,\lQ)$ that has a minimum in $r= r_0\sim 1/\lQ$ ($r=\vert {\bf r}\vert$, {\bf r} being the relative coordinate) . Hence the energy of the small fluctuation about that minimum
is $E \sim \sqrt{\lQ^3/m_Q} \ll \lQ \ll m_Q$. Consequently, we are in a situation analogous to the strong coupling regime of pNRQCD in which the scale $\lQ$ may be integrated out. It then makes sense to restrict the study to the lowest lying heavy hybrid potentials, $\Sigma_u^-$ and $\Pi_u$, since the gap to the next state is parametrically $O(\lQ)$.

The potentials associated with $\Sigma_u^-$ and $\Pi_u$ are degenerated at short distances (see fig. \ref{fig2}). In pNRQCD this is easily understood as they correspond to different projections with respect to {\bf r} of the same operator $tr({\bf B}({\bf 0}, t) {\rm O}({\bf 0},{\bf r}, t))$ (we have set the center of mass coordinate {\bf R}={\bf 0}, see \cite{Brambilla:1999xf} for the notation). It is then natural to associate to the lower lying heavy hybrids a vectorial wave function ${\bf H}({\bf 0},{\bf r}, t)$ with the same symmetry properties as the above operator, such that its projection to {\bf r} evolves with $V_{\Sigma_u^-}$ and its projection orthogonal to {\bf r} with $V_{\Pi_u}$. We then have the following Lagrangian density,

\be
\label{h}
{\cal L}={\rm tr} \left( {H^i}^\dagger\left( \delta_{ij}i\partial_0-{h_H}_{ij}\right)H^j \right) \quad ,\quad
{h_{H}}_{ij}\!=\!\left(-\frac{\nabla^2}{m_Q}\!+\!V_{\Sigma_u^-}(r)\right)\delta_{ij}\!+\!\left(\delta_{ij}\!-\!\hat{r}_{i} \hat{r}_{j}\right)\left[V_{\Pi_u}(r)\!-\!V_{\Sigma_u^-}(r)\right]\,,
\ee
${\bf \hat r}={\bf r}/\vert {\bf r}\vert$.
${\bf H}={\bf H}({\bf 0},{\bf r}, t)$ is a matrix in spin space and the trace is over spin indices.
 ${h_H}_{ij}$ above does not depend on the spin of the quarks, and hence it is invariant under spin symmetry transformations, but it does depend on the total angular momentum of the gluonic degrees of freedom ${\bf L}_g$, in this case $L_g=1$ as it is apparent from the vectorial character of {\bf H}.
The $P$ and $C$ associated to a Hybrid state with quark-antiquark orbital angular momentum $L$ and quark-antiquark spin $S$
become,
\be
P=(-1)^{L+1} \; ,\, C=(-1)^{L+S+1}\,.
\ee  
Leaving aside the spin of the quarks, it is convenient to express {\bf H} in a basis of eigenfunctions of ${\bf J}= {\bf L} + {\bf L}_g$, where {\bf L} is the orbital angular momentum of the quarks. This is achieved
using  Vector Spherical Harmonics \cite{pg},
\be
\label{jJML}
{\bf H}({\bf 0},{\bf r},t)=\frac{e^{-iEt}}{r}\left(P_0^+(r)\mathcal{\bf Y}_{00}^{L=1}(\theta,\phi) +\sum_{J=1}^{\infty}\sum_{M=-J}^{J}[P_J^{+}(r)\mathcal{\bf Y}_{JM}^{L=J+1}(\theta,\phi)
 +P_J^{0}(r)\mathcal{\bf Y}_{JM}^{L=J}(\theta,\phi)+P_J^{-}(r)\mathcal{\bf Y}_{JM}^{L=|J-1|}(\theta,\phi)]\right)\,.
\ee
Note that ${\bf J}$ is a conserved quantity thanks to heavy quark spin symmetry.
The eigenvalue problem then reduces for $J\not=0$ to
\be
\label{cop}
	\left[ 
 -\frac{1}{m_Q}\frac{\partial^2}{\partial r^2}\!+
\begin{pmatrix}
 		\frac{(J-1)J}{m_Qr^2}\!  & 0                    \\
 		0 & 	\frac{(J+1)(J+2)}{m_Qr^2}\!                \\
 	\end{pmatrix} +
\!V_{\Sigma_u^-}(r) +
	V_q(r) \begin{pmatrix}
 		\!\frac{J+1}{2J+1}  & \frac{\sqrt{(J+1)J}}{2J+1}                    \\
 		\frac{\sqrt{(J+1)J}}{2J+1} & 	\frac{J}{2J+1}                    \\
 	\end{pmatrix}
 	\!
 	\right]\! 
	\begin{pmatrix}
 		P_J^{-}(r) \\
 		P_J^{+}(r) \\
 	\end{pmatrix}\!
	\nn
 	=\!E\!
 	\begin{pmatrix} 
 		P_J^{-}(r) \\
 		P_J^{+}(r) \\
 	\end{pmatrix} 
	\ee

	\be
	\left(\!-\frac{1}{m_Q}\frac{\partial^2}{\partial r^2}\!+\!\frac{J(J+1)}{m_Qr^2}\!+\!V_{\Pi_u}(r)\right)P_J^{0}(r)\!=\!EP_J^{0}(r) \,,
	\label{uncop}
	\ee
	where $V_q(r)=V_{\Pi_u}(r)-V_{\Sigma_u^-}(r)$. For $J=0$ only the equation for $P_J^{+}(r) $ remains.
	
We are going to approximate $V_{\Pi_u}(r)$ and $V_{\Sigma_u^-}(r)$ by simple functions that have the correct asymptotic behavior at short and long distances, and fit well the lattice data of ref. \cite{Juge:2002br}.  For $V_{\Sigma_u^-}(r)$ it is enough to take a Cornell-like potential 
in order to get a good description of lattice data. We impose the linear term to be equal to the one for heavy quarkonium ($V_{\Sigma_g^+}(r)$)  as dictated by the QCD effective string theory (EST) \cite{Luscher:2004ib}, and the Coulomb term to be related to the one of $V_{\Sigma_g^+}(r)$ as dictated by perturbation theory. Hence, only an additive constant is left as a free parameter,
 which can in turn be related to the corresponding constant for $V_{\Sigma_g^+}(r)$ 
through the lattice data.  
The last constant is fixed 
by fitting the heavy quarkonium spectrum. 
For $V_{\Pi_u}(r)$ a Cornell-like form does not fit lattice data well at intermediate distances. Hence, we take a slightly more complicated function for it, we substitute the Coulomb term by a rational function with five free parameters. At short and long distances we impose $V_{\Pi_u}(r)-V_{\Sigma_u^-}(r)$ to be compatible with the weak coupling pNRQCD results at NLO \cite{Brambilla:1999xf}
and with the EST results at NLO \cite{Luscher:2004ib} respectively. The remaining two free parameters are fitted to lattice data.
 
Using the potentials above as an input we solve numerically (\ref{cop}) and obtain the results displayed in tables \ref{cEspectrum} and \ref{bEspectrum}. 
We also display the results for heavy quarkonium, obtained with a Cornell potential that fits well the lattice data for $\Sigma_g^+$ of \cite{Juge:2002br}.

\begin{table}[htbp]
	\centering
	\begin{tabular}{|c|c|c|c|c|c|c|}
		\hline
		&     &            &        &      $S=0 $     &      $S=1$      &                             \\
		$NL_J$ & w-f & $c\bar{c}$ & Hybrid & $\mathcal{J}^{PC}$ & $\mathcal{J}^{PC}$ & $\Lambda^{\epsilon}_{\eta}$ \\ \hline \hline
		$1s$ & S & 3068 &  & $0^{-+}$ & $1^{--}$ & $\Sigma_g^+$ \\ 
		$2s$ & S & 3678 &  & $0^{-+}$ & $1^{--}$ & $\Sigma_g^+$ \\ 
		$3s$ & S & 4131 &  & $0^{-+}$ & $1^{--}$ & $\Sigma_g^+$ \\ 
		$1p_0$ & H$^+$ &  & 4486 & $0^{++}$ & $1^{+-}$ & $\Sigma_u^-$ \\ 
		$4s$ & S & 4512 &  & $0^{-+}$ & $1^{--}$ & $\Sigma_g^+$ \\ 
		$2p_0$ & H$^+$ &  & 4920 & $0^{++}$ & $1^{+-}$ & $\Sigma_u^-$ \\ 
		$3p_0$ & H$^+$ &  & 5299 & $0^{++}$ & $1^{+-}$ & $\Sigma_u^-$ \\ 
		$4p_0$ & H$^+$ &  & 5642 & $0^{++}$ & $1^{+-}$ & $\Sigma_u^-$ \\ \hline
		$1p$ & S & 3494 &  & $1^{+-}$ & $(0,1,2)^{++}$ & $\Sigma_g^+$ \\ 
		$2p$ & S & 3968 &  & $1^{+-}$ & $(0,1,2)^{++}$ & $\Sigma_g^+$ \\ 
		$1(s/d)_1$ & H$^{+-}$ &  & 4011 & $1^{--}$ & $(0,1,2)^{-+}$ & $\Pi_u\Sigma_u^-$ \\ 
		$1p_1$ & H$^0$ &  & 4145 & $1^{++}$ & $(0,1,2)^{+-}$ & $\Pi_u$ \\ 
		$2(s/d)_1$ & H$^{+-}$ &  & 4355 & $1^{--}$ & $(0,1,2)^{-+}$ & $\Pi_u\Sigma_u^-$ \\ 
		$3p$ & S & 4369 &  & $1^{+-}$ & $(0,1,2)^{++}$ & $\Sigma_g^+$ \\
		$2p_1$ & H$^0$ &  & 4511 & $1^{++}$ & $(0,1,2)^{+-}$ & $\Pi_u$ \\ 
		$3(s/d)_1$ & H$^{+-}$ &  & 4692 & $1^{--}$ & $(0,1,2)^{-+}$ & $\Pi_u\Sigma_u^-$ \\ 
		$4(s/d)_1$ & H$^{+-}$ &  & 4718 & $1^{--}$ & $(0,1,2)^{-+}$ & $\Pi_u\Sigma_u^-$ \\ 
		$4p$ & S & 4727 &  & $1^{+-}$ & $(0,1,2)^{++}$ & $\Sigma_g^+$ \\ 
		$3p_1$ & H$^0$ &  & 4863 & $1^{++}$ & $(0,1,2)^{+-}$ & $\Pi_u$ \\ 
		$5(s/d)_1$ & H$^{+-}$ &  & 5043 & $1^{--}$ & $(0,1,2)^{-+}$ & $\Pi_u\Sigma_u^-$ \\ 
		$5p$ & S & 5055 &  & $1^{+-}$ & $(0,1,2)^{++}$ & $\Sigma_g^+$ \\ \hline
		$1d$ & S & 3793 &  & $2^{-+}$ & $(1,2,3)^{--}$ & $\Sigma_g^+$ \\ 
		$2d$ & S & 4210 &  & $2^{-+}$ & $(1,2,3)^{--}$ & $\Sigma_g^+$ \\ 
		$1(p/f)_2$ & H$^{+-}$ &  & 4231 & $2^{++}$ & $(1,2,3)^{+-}$ & $\Pi_u\Sigma_u^-$ \\ 
		$1d_2$ & H$^0$ &  & 4334 & $2^{--}$ & $(1,2,3)^{-+}$ & $\Pi_u$ \\ 
		$2(p/f)_2$ & H$^{+-}$ &  & 4563 & $2^{++}$ & $(1,2,3)^{+-}$ & $\Pi_u\Sigma_u^-$ \\ 
		$3d$ & S & 4579 &  & $2^{-+}$ & $(1,2,3)^{--}$ & $\Sigma_g^+$ \\ 
		$2d_2$ & H$^0$ &  & 4693 & $2^{--}$ & $(1,2,3)^{-+}$ & $\Pi_u$ \\ 
		$3(p/f)_2$ & H$^{+-}$ &  & 4886 & $2^{++}$ & $(1,2,3)^{+-}$ & $\Pi_u\Sigma_u^-$ \\ 
		$4d$ & S & 4916 &  & $2^{-+}$ & $(1,2,3)^{--}$ & $\Sigma_g^+$ \\ 
		$4(p/f)_2$ & H$^{+-}$ &  & 4923 & $2^{++}$ & $(1,2,3)^{+-}$ & $\Pi_u\Sigma_u^-$ \\ 
		$3d_2$ & H$^0$ &  & 5036 & $2^{--}$ & $(1,2,3)^{-+}$ & $\Pi_u$ \\ \hline
	\end{tabular}
	\caption{ Charmonium (S) and Hybrid charmonium (H$^{+-0}$)  energy spectrum computed with 
	$m_c=1.47GeV$. Masses are in MeV. States which only differ by the heavy quark spin $(S=0,1)$ are degenerated.
	$N$ is the principal quantum number, $L$ the orbital angular momentum of the heavy quarks, $J$ is $L$ plus the total angular momentum of the gluons, $S$ the spin of the heavy quarks and ${\cal J}$ is the total angular momentum. 
	For quarkonium, $J$ coincides with $L$ and it is not displayed.
	The last column shows the relevant potentials for each state. The $(s/d)_1$, $p_1$ and $p_0$ states are usually named $H_1$, $H_2$ and $H_3$ respectively. 
	}
	\label{cEspectrum}
\end{table}

	\begin{table}[htbp]
		\centering
		\begin{tabular}{|c|c|c|c|c|c|c|}
			\hline 
			&     &            &        &  $S=0$ & $S=1$ & \\
			$NL_J$ & w-f & $b\bar{b}$ & Hybrid & $\mathcal{J}^{PC}$ & $\mathcal{J}^{PC}$ & $\Lambda^{\epsilon}_{\eta}$ \\ \hline \hline
			$1s$ & S & 9442 &  & $0^{-+}$ & $1^{--}$ & $\Sigma_g^+$ \\ 
			$2s$ & S & 10009 &  & $0^{-+}$ & $1^{--}$ & $\Sigma_g^+$ \\ 
			$3s$ & S & 10356 &  & $0^{-+}$ & $1^{--}$ & $\Sigma_g^+$ \\ 
			$4s$ & S & 10638 &  & $0^{-+}$ & $1^{--}$ & $\Sigma_g^+$ \\ 
			$1p_0$ & H$^+$ &  & 11011 & $0^{++}$ & $1^{+-}$ & $\Sigma_u^-$ \\ 
			$2p_0$ & H$^+$ &  & 11299 & $0^{++}$ & $1^{+-}$ & $\Sigma_u^-$ \\ 
			$3p_0$ & H$^+$ &  & 11551 & $0^{++}$ & $1^{+-}$ & $\Sigma_u^-$ \\ 
			$4p_0$ & H$^+$ &  & 11779 & $0^{++}$ & $1^{+-}$ & $\Sigma_u^-$ \\ \hline
			$1p$ & S & 9908 &  & $1^{+-}$ & $(0,1,2)^{++}$ & $\Sigma_g^+$ \\ 
			$2p$ & S & 10265 &  & $1^{+-}$ & $(0,1,2)^{++}$ & $\Sigma_g^+$ \\ 
			$3p$ & S & 10553 &  & $1^{+-}$ & $(0,1,2)^{++}$ & $\Sigma_g^+$ \\ 
			$1(s/d)_1$ & H$^{+-}$ &  & 10690 & $1^{--}$ & $(0,1,2)^{-+}$ & $\Pi_u\Sigma_u^-$ \\ 
			$1p_1$ & H$^0$ &  & 10761 & $1^{++}$ & $(0,1,2)^{+-}$ & $\Pi_u$ \\ 
			$4p$ & S & 10806 &  & $1^{+-}$ & $(0,1,2)^{++}$ & $\Sigma_g^+$ \\ 
			$2(s/d)_1$ & H$^{+-}$ &  & 10885 & $1^{--}$ & $(0,1,2)^{-+}$ & $\Pi_u\Sigma_u^-$ \\ 
			$2p_1$ & H$^0$ &  & 10970 & $1^{++}$ & $(0,1,2)^{+-}$ & $\Pi_u$ \\ 
			$5p$ & S & 11035 &  & $1^{+-}$ & $(0,1,2)^{++}$ & $\Sigma_g^+$ \\ 
			$3(s/d)_1$ & H$^{+-}$ &  & 11084 & $1^{--}$ & $(0,1,2)^{-+}$ & $\Pi_u\Sigma_u^-$ \\ 
			$4(s/d)_1$ & H$^{+-}$ &  & 11156 & $1^{--}$ & $(0,1,2)^{-+}$ & $\Pi_u\Sigma_u^-$ \\ 
			$3p_1$ & H$^0$ &  & 11175 & $1^{++}$ & $(0,1,2)^{+-}$ & $\Pi_u$ \\
			$6p$ & S & 11247 &  & $1^{+-}$ & $(0,1,2)^{++}$ & $\Sigma_g^+$ \\ 
			$5(s/d)_1$ & H$^{+-}$ &  & 11284 & $1^{--}$ & $(0,1,2)^{-+}$ & $\Pi_u\Sigma_u^-$ \\ \hline
			$1d$ & S & 10155 &  & $2^{-+}$ & $(1,2,3)^{--}$ & $\Sigma_g^+$ \\ 
			$2d$ & S & 10454 &  & $2^{-+}$ & $(1,2,3)^{--}$ & $\Sigma_g^+$ \\ 
			$3d$ & S & 10712 &  & $2^{-+}$ & $(1,2,3)^{--}$ & $\Sigma_g^+$ \\
			$1(p/f)_2$ & H$^{+-}$ &  & 10819 & $2^{++}$ & $(1,2,3)^{+-}$ & $\Pi_u\Sigma_u^-$ \\ 
			$1d_2$ & H$^0$ &  & 10870 & $2^{--}$ & $(1,2,3)^{-+}$ & $\Pi_u$ \\ 
			$4d$ & S & 10947 &  & $2^{-+}$ & $(1,2,3)^{--}$ & $\Sigma_g^+$ \\ 
			$2(p/f)_2$ & H$^{+-}$ &  & 11005 & $2^{++}$ & $(1,2,3)^{+-}$ & $\Pi_u\Sigma_u^-$ \\ 
			$2d_2$ & H$^0$ &  & 11074 & $2^{--}$ & $(1,2,3)^{-+}$ & $\Pi_u$ \\ 
			$5d$ & S & 11163 &  & $2^{-+}$ & $(1,2,3)^{--}$ & $\Sigma_g^+$ \\ 
			$3(p/f)_2$ & H$^{+-}$ &  & 11197 & $2^{++}$ & $(1,2,3)^{+-}$ & $\Pi_u\Sigma_u^-$ \\ 
			$3d_2$ & H$^0$ &  & 11275 & $2^{--}$ & $(1,2,3)^{-+}$ & $\Pi_u$ \\ 
			$4(p/f)_2$ & H$^{+-}$ &  & 11291 & $2^{++}$ & $(1,2,3)^{+-}$ & $\Pi_u\Sigma_u^-$ \\ \hline
		\end{tabular}
		\caption{ Bottomonium (S) and hybrid bottomonium (H$^{+-0}$)  energy spectrum computed with $m_b=4.88GeV$. 
		Masses are in MeV. States which only differ by the heavy quark spin $(S=0,1)$ are degenerated.
	$N$ is the principal quantum number, $L$ the orbital angular momentum of the heavy quarks, $J$ is $L$ plus the total angular momentum of the gluons, $S$ the spin of the heavy quarks and ${\cal J}$ is the total angular momentum. 
	For quarkonium, $J$ coincides with $L$ and it is not displayed.
	The last column shows the relevant potentials for each state. The $(s/d)_1$, $p_1$ and $p_0$ states are usually named $H_1$, $H_2$ and $H_3$ respectively. 
	}
		\label{bEspectrum}
	\end{table}

In table \ref{id} we show possible identifications with XYZ states. According to this table only spin zero hybrids would have been observed. It is interesting to notice that $Y(4008)$, $Y(4360)$ and $Y(4660)$ would correspond to the ground state and the lower excitations of the $(s/d)_1$ state. However, the three states have been observed to decay to vector quarkonium, which violates
spin symmetry \cite{Braaten:2014ita}. This is also so for $Y_b(10890)$. In fact, from the $1^{--}$ candidates only for $X(4630)$ no spin symmetry violating decay has been observed. This motivates the study of mixing with heavy quarkonium in section \ref{mixing}. We report in the next section on our results for the decay widths to lower lying heavy quarkonium states.   

\begin{table}[htbp]
\centering
	\begin{tabular}{|c|c|c|c|c|c|c|}
		\hline
		State & M & $J^{PC}$ & XYZ  & M$_{exp}$ & $\Gamma_{exp}$ & $J^{PC}_{exp}$ \\ \hline
		$1(s/d)_1$ & 
		4011 & $1^{--}$,$(0,1,2)^{-+}$ & Y(4008) & $4008^{+121}_{-49}$ & $226\pm97$ & $1^{--}$ \\ \hline
		$1p_1$ & 
		4145 & $1^{++}$,$(0,1,2)^{+-}$ & Y(4140) & $4144.5\pm2.6$ & $15^{+11}_{-7}$ & $?^{?+}$ \\ 
		\multicolumn{1}{|l|}{} & \multicolumn{1}{l|}{} & \multicolumn{1}{l|}{} & X(4160) & $4156^{+29}_{-25}$ & $139^{+113}_{-65}$ & $?^{?+}$ \\ \hline
		$2(s/d)_1$ & 
		4355 & $1^{--}$,$(0,1,2)^{-+}$ & Y(4360) & $4361\pm13$ & $74\pm18$ & $1^{--}$ \\ \hline
		$3(s/d)_1$ & 
		4692 & $1^{--}$,$(0,1,2)^{-+}$ & Y(4660) & $4664\pm12$ & $48\pm15$ & $1^{--}$ \\ 
		\multicolumn{1}{|l|}{} & \multicolumn{1}{l|}{} & \multicolumn{1}{l|}{} & X(4630) & $4634^{+9}_{-11}$ & $92^{+41}_{-32}$ & $1^{--}$ \\
		\hline \hline
		$2(s/d)_1$ & 
		10885 & $1^{--}$,$(0,1,2)^{-+}$ & $Y_b$(10890) & $10888.4\pm3$ & $30.7^{+8.9}_{-7.7}$ & $1^{--}$ \\ \hline
	\end{tabular}
\caption{Possible identifications of XYZ states in our spectrum. Masses and decay widths are given in MeV.}
	\hfill
\label{id}
	\end{table}

\section{Decay}
\label{decay}

Since we are interested in the lowest lying heavy hybrid states, it is enough for us to consider an effective theory  for energy fluctuations $E\ll \lQ$ around those states. The energy gap to the lower lying heavy quarkonium states is greater than $\lQ$. Hence the lower lying heavy quarkonium states can be integrated out, which will give rise to an imaginary potential $\Delta V$ for the heavy hybrid states, which in turn will produce a decay width for them, 
$\Gamma_{H\to S}=-2$ $\langle H|\text{Im}\Delta V|H\rangle$. This is much in the same way as integrating out hard gluons produces operators with imaginary matching coefficients in NRQCD \cite{Bodwin:1994jh}. Furthermore, if we assume that the energy gap $\Delta E$ fulfills  $\Delta E\gg \lQ$, and that the process is dominated by short distances, the integration can be done using the weak coupling regime of pNRQCD \cite{Pineda:1997bj,Brambilla:1999xf}

We obtain,
\begin{equation}
	{\rm Im}\Delta V=-
	\frac{2}{3}\frac{\als T_F}{N_c} \sum_n r^i|S_n\rangle\langle S_n|r^i\, 
	(i\partial_t-E_n)^3\,,
	\end{equation} 
$T_F=1/2$, $N_c=3$, and $\als$ is the QCD strong coupling constant. $E_n$ is the energy  
of the n-th heavy quarkonium state, $S_n$. 
The calculation is reliable only for those states that fulfill $\Delta E_n\gg \lQ$, for which we identify,
\begin{equation}
	\Gamma(H_m\!\to \!S_n)\!=\!
	\!\frac{4}{3}\!\frac{\als T_F}{ N_c}\! 
\langle H_m|r^i|S_n
\rangle \!
\langle S_n|r^i|H_m
\rangle \!(\Delta E_{mn})^3\, ,
	\end{equation} 
where $m$ stands for $NL_J$, the quantum numbers of the heavy hybrid ($H_m$), $n$ for $N'L'$, the quantum numbers of the heavy quarkonium ($S_n$), and $\Delta 
E_{mn}$ is the energy difference between them.
For consistency, $\left\langle S_n|r^i|H_m\right\rangle \!\Delta E_{mn}$ should also be small, otherwise the multipole expansion built in weak coupling pNRQCD would not be justified.
	The structure of the decay width above implies that no heavy hybrid with $L=J$ decays to heavy quarkonium at this order. This selects $X(4160)$ as the preferred candidate for the $1p_1$ state in table \ref{id} since no decay to charmonium has been observed, as opposite to $X(4140)$.
The numerical values of the decay widths are given in table \ref{decayW}.  
The scale of $\als$ is taken at $\mu=\Delta E_{mn}$. At this order, the decays respect heavy quark spin symmetry, and hence the spin of the heavy quarks must be the same in the initial heavy hybrid state and in the 
final heavy quarkonium state. For the states in table \ref{id}, we are only able to put forward a couple of reliable estimates
\be
 \Gamma (Y(4360)\to h_c + X) \sim 19 \,{\rm MeV} \quad ,\quad
\Gamma (Y_b(10890)\to h_b + X) \sim 9.6 \,{\rm MeV}\,.
\ee
 Notice that both numbers above are smaller than the measured total decay widths, as they should.

\begin{table}[htbp]
	\centering
	\begin{tabular}{|c|c|c|c|c|}
		\hline

			$NL_J \rightarrow N'L'$ & {$\Delta E$ (MeV)} & {$\langle r\rangle _{NL_J,\,N'L'}$ (GeV$^{-1}$)} & ${\vert\Delta E \langle r\rangle _{NL_J,\,N'L'}\vert}$ & {$\Gamma$ (MeV) }  \\ \hline
					
		$1p_0\rightarrow 2s$ & {808} & {0.40} & {0.32} & {6.1} \\ 
		$2(s/d)_1\rightarrow 1p$ & 861 & 0.63 & 0.54 & 19 \\ 
		\hline
		\hline
		$1p_0\rightarrow 1s$ & {1569} & {-0.416} & {0.65} & {31}   \\ 
		$1p_0\rightarrow 2s$ & {1002} & {0.432} & {0.43} & {8.7}   \\  
		$2p_0\rightarrow 2s$ & {1290} & {-0.137} & {0.18} & {1.9}   \\ 
		$2p_0\rightarrow 3s$ & {943} & {0.462} & {0.44} & {8.3}   \\ 
		{$2(s/d)_1\rightarrow 1p$} & {977} & {0.470} & {0.46} & {9.6}  \\ 
		\hline
	\end{tabular}
	\caption{Decay widths for hybrid charmonium (above) and bottomonium (below) to lower lying charmonia and bottomonia respectively. We only display results
	for which $\Delta E \gtrsim 800$MeV and ${\vert\Delta E \langle r\rangle _{NL_J\,,N'L'}\vert}\lesssim 0.6$.}\,.
	\label{decayW}
\end{table}	
	
	\section{Mixing}
\label{mixing}

		We shall focus here 
		on the mixing with heavy quarkonium, basically because it is amenable to a systematic treatment. The mixing of spin zero heavy hybrids with spin one heavy quarkonium is relevant because it  
		may explain the spin symmetry violating decays observed in some hybrid candidates.  
		In the static limit, heavy quarkonium 
		and heavy hybrids 
		do not mix by construction (they are built as orthogonal states). Hence, the mixing must be due to $1/m_Q$ corrections to 
the BO approximation. These corrections may become large if there are energy levels of heavy quarkonium close to those of heavy hybrids. A way to systematically compute $1/m_Q$ corrections for heavy quarkonium was established in \cite{Brambilla:2000gk,Pineda:2000sz} for the strong coupling regime of pNRQCD.  
The formalism in \cite{Brambilla:2000gk} can also be used to calculate the mixing potentials as we sketch below.
We may generally consider an effective theory for energy fluctuations $E$ around a hybrid state, such that $E\ll \lQ$. If there is
a heavy quarkonium state close to that energy, we may expect it to modify the value of the energy $E$. This effective theory reads, 	
\begin{equation}
	\label{H+S}
	\mathcal{L}_{H+S}=
	   {\rm tr}\left( S^\dagger [i\partial_0-h_s]S)\right) + {\rm tr}\left( H^{i\dagger}[i\delta_{ij}\partial_0-{h_H}_{ij}]H^j\right)
		+ {\rm tr}\left(S^\dagger  V_S^{ij} \left\{ \sigma^i \, , H^j\right\}\, + {\rm H.c.}\right)\,.
	\end{equation}
	The traces are over spin indices and 
	\be
	\label{mixdecomp}
	V_S^{ij}=V_S^{ij}({\bf r})=\delta^{ij}V_S^\Pi(r) +{\hat r}^i{\hat r}^j(V_S^\Sigma(r)-V_S^\Pi(r))
	\ee
	is the mixing potential, $h_s=-\frac{\nabla^2}{m_Q}\!+\!V_{\Sigma_g^+}(r)$ and ${h_H}_{ij}$ is made explicit in (\ref{h}). Notice that the last term in (\ref{H+S}) mixes spin zero (one) heavy hybrids with spin one (zero) heavy quarkonium.
	
	By matching NRQCD with the effective theory above at order $1/m_Q$ we obtain,
	\be
\label{lattice}
\frac{\frac{c_F}{2m_Q}\int_{-\frac{T}{2}}^{\frac{T}{2}} dt <{\bf \hat r}g{\bf B}(\frac{\bf r}{2},t)\, {\bf \hat r}g{\bf B}({\bf 0},-\frac{T}{2})>_\Box}
{<1>_\Box^{1/2}<{\bf \hat r}g{\bf B}({\bf 0},\frac{T}{2})\, {\bf \hat r}g{\bf B}({\bf 0},-\frac{T}{2})>_\Box^{1/2}} 
=2V_S^\Sigma \frac{\sin \left( \frac{(V_{\Sigma_u^-}-V_{\Sigma_g^+}) T}{2}\right)}{ V_{\Sigma_u^-}-V_{\Sigma_g^+}}
\ee
\be
\frac{\frac{c_F}{2m_Q}\int_{-\frac{T}{2}}^{\frac{T}{2}} dt <g{\bf B}(\frac{\bf r}{2},t)g{\bf B}({\bf 0},-\frac{T}{2})-{\bf \hat r}g{\bf B}(\frac{\bf r}{2},t)\, {\bf \hat r}g{\bf B}({\bf 0},-\frac{T}{2})>_\Box}
{<1>_\Box^{1/2} <g{\bf B}({\bf 0},\frac{T}{2})g{\bf B}({\bf 0},-\frac{T}{2})-{\bf \hat r}g{\bf B}({\bf 0},\frac{T}{2})\, {\bf \hat r}g{\bf B}({\bf 0},-\frac{T}{2})>_\Box^{1/2}} 
=2\sqrt{2}V_S^\Pi \frac{\sin \left( \frac{(V_{\Pi_u}-V_{\Sigma_g^+}) T}{2}\right)}{ V_{\Pi_u}-V_{\Sigma_g^+}}\,,
\ee
$<\dots>_\Box$ means expectation value in the Wilson loop, and $c_F$ is an NRQCD hard matching coefficient.
Notice that the Euclidean version of the objects on the lhs can be easily calculated on the lattice. At large $T$, $V_S^\Sigma$ and $V_S^\Pi$ can be then extracted by matching the data to the Euclidean version of the rhs, 
since $V_{\Sigma_g^+}$, $V_{\Sigma_u^-}$ and $V_{\Pi_u}$ are known. In the following we are going to derive short and long distance constraints on these potentials using weak coupling pNRQCD \cite{Pineda:1997bj,Brambilla:1999xf} and the EST  
respectively \cite{Luscher:2002qv,Luscher:2004ib}.

At short distances the time evolution of a $Q\bar Q$ pair is described by pNRQCD at weak coupling \cite{Pineda:1997bj,Brambilla:1999xf}. 
The leading spin-dependent term in the pNRQCD Lagrangian reads \cite{Brambilla:2002nu},
\be
\label{pNRQCD'}
	{\mathcal{L}}_{pNRQCD}={c_F \over 2m_Q} {\rm Tr} \left( {\rm O}^\dagger({\bf r}, {\bf R}, t)\, g{\bf B}({\bf R}, t)\,\{ \boldsymbol{\sigma} , {\rm S}({\bf r}, {\bf R}, t)\} \right)
	+ \hbox{H.c.}\, .
	\ee
We use $tr$ for trace over color indices and Tr for
 trace over both color and spin indices. Notice that the term above shows an
{\bf r}-independent interaction between	the single field and the operator ${ tr} ({\rm O}\,{\bf B})$, the short distance representation of {\bf H}, which implies that
\be
V_S^\Sigma(r)=V_S^\Pi(r)= \pm\frac{c_F \lambda^2}{m_Q}\, ,
\label{sdc}
\ee
where we have made the sign explicit, and $\lambda\sim \lQ$ is a constant.

At long distances the energy spectrum of a static $Q\bar Q$ pair is well described by the 
EST \cite{Luscher:2002qv,Luscher:2004ib}. The mapping between operator insertions in the temporal Wilson lines of the 
Wilson loop and the corresponding operators in the EST was established in \cite{PerezNadal:2008vm}. A mapping can also be established for operators in the space
Wilson lines at $\pm T/2$ and states in the EST \cite{os}. Hence, the expressions on the lhs of (\ref{lattice}) can be calculated in the EST (as well as $V_{\Sigma_u^-}$, $V_{\Pi_u}$ and $V_{\Sigma_g^+}$). We obtain, 
\be
V_S^\Sigma(r)=-\frac{\pi^2\Lambda'''c_F}{m_Q\kappa r^3}\quad ,\quad V_S^\Pi(r)=\frac{\pi^{3/2}\Lambda'c_F}{2m_Q\sqrt{\kappa} r^2}\, .
\ee
The parameters $\Lambda'\sim \lQ$ and $\Lambda'''\sim \lQ$ also appear in the spin-orbit  and tensor potentials of heavy quarkonium \cite{PerezNadal:2008vm,Brambilla:2014eaa} , which have been calculated on the lattice. We obtain from fits to the data of ref. \cite{Koma:2009ws},
\be
\kappa\sim 0.187 {\rm GeV^2}\quad ,\quad \Lambda'\sim -118 {\rm MeV} \quad ,\quad \Lambda'''\sim \pm 230 {\rm MeV}\,,
\ee
where we have also displayed the value we use for the string tension $\kappa$. 

For the actual mixing potentials, we use the simplest interpolation that allows for a sign flip between the short and long distance expressions without introducing any further scale. We have explored the following values for the only unknown parameter $\lambda= 100, 300, 600$ MeV, and all possible sign combinations. We present here our results for the phenomenologically relevant candidates to charmonium hybrids in table \ref{id} ($S=0$). For the $1^{++}$ and $1^{--}$ states a system of two and four coupled equations must be solved respectively.

We find that for the $1^{++}$ states, the mixing is small in all cases ($< 5\%$). For the $1^{--}$ states this is only so in the case that $V_S^\Sigma$ flips sign and $V_S^\Pi$ does not ($< 6\%$). In the cases that both $V_S^\Sigma$ and $V_S^\Pi$ flip sign or both keep the same sign, it may reach $15\%$ for some states when $\lambda=600$ MeV. The most dramatic case occurs when $V_S^\Sigma$ keeps the same sign and $V_S^\Pi$ flips it, the mixing ranges from $20\%$ to $40\%$
for all the states in table \ref{id} when $\lambda=600$ MeV. In this case the charmonium hybrid spectrum moves up about $10-20$ MeV.

\section{Conclusions}

We have calculated the lower lying spectrum for charmonium and bottomonium hybrids at leading order in the BO approximation, including the $\Pi_u$-$\Sigma_u^-$ mixing terms, which were neglected in most of the earlier calculations \cite{Juge:1999ie,Braaten:2014qka}. A detailed comparison with earlier work will be presented elsewhere \cite{os}. Let us only mention here that the equations we get (\ref{cop})  
are equivalent to those obtained in \cite{Berwein:2015vca} (see also \cite{jt}), and our energy spectrum for $H_1$ and $H_2$ is at the lower end of the error bars in that reference whereas for $H_3$ it is in the middle. The basic difference with respect to ref. \cite{Berwein:2015vca} is that we fix the arbitrary constant in the potential through the heavy quarkonium spectrum whereas in that reference it is fixed through the RS heavy quark masses \cite{Pineda:2001zq}. We identify a number of possible heavy hybrid states in table \ref {id}, which would correspond to spin zero hybrids. The $1^{--}$ states would be the ground state and lower excitation of the $(s/d)_1$ ($H_1$) heavy hybrid states. However, spin symmetry violating decays have been observed for all of them, except for $X(4630)$.

Using weak coupling pNRQCD, we have also been able to estimate a limited number of decay widths to heavy quarkonium. In the way, we have learned that heavy hybrids with $J=L$ do not decay to lower lying heavy quarkonium states at leading order. This selects $X(4160)$ as a good candidate for the spin zero $1p_1$ ($H_2$) hybrid state.

We have pointed out that spin zero (one) heavy hybrids mix with spin one (zero) heavy quarkonium. We have calculated the asymptotic behavior at short and long distances for the mixing potentials, using weak coupling pNRQCD and the EST respectively. By building simple interpolations, we have also demonstrated that the mixing may be sizable ($20\%-40\%$) for the $1^{--}$ charmonium hybrid states, which may explain the observed decays to spin one charmonium. As a consequence, $Y(4008)$, $Y(4360)$ and $Y(4660)$ remain as spin zero $(s/d)_1$ ($H_1$) heavy hybrid candidates.  

\vspace{0.5cm}

{\bf Acknowledgments:} We have been supported by the  Spanish Excellence Network on
Hadronic Physics FIS2014-57026-REDT.  J.S. also acknowledges support from
the 2014-SGR-104 grant (Catalonia), the FPA2013-4657 and FPA2013-43425-P projects, and the CPAN CSD2007-00042 Consolider–Ingenio 2010 program (Spain).

\end{document}